\documentclass{ws-procs9x6}                             
\usepackage{epsfig}       
\usepackage{amssymb}     
\usepackage{amsfonts}    
\newskip\humongous \humongous=0pt plus 1000pt minus 1000pt

\newif\ifdtup

\catcode`\@=11
 
\@addtoreset{equation}{section}
\def\theequation{{\thesection}.\arabic{equation}}
 
\def\@normalsize{\@setsize\normalsize{15pt}\xiipt\@xiipt
\abovedisplayskip 14pt plus3pt minus3pt%
\belowdisplayskip \abovedisplayskip
\abovedisplayshortskip \z@ plus3pt%
\belowdisplayshortskip 7pt plus3.5pt minus0pt}
 
\def\small{\@setsize\small{13.6pt}\xipt\@xipt
\abovedisplayskip 13pt plus3pt minus3pt%
\belowdisplayskip \abovedisplayskip
\abovedisplayshortskip \z@ plus3pt%
\belowdisplayshortskip 7pt plus3.5pt minus0pt
\def\@listi{\parsep 4.5pt plus 2pt minus 1pt
     \itemsep \parsep
     \topsep 9pt plus 3pt minus 3pt}}
 
\relax

\catcode`@=12
 
\catcode`\@=11
 
\def\section{\@startsection{section}{1}{\z@}{3.5ex plus 1ex minus
   .2ex}{2.3ex plus .2ex}{\large\bf}}
 
\def\thesection{\arabic{section}}
\def\thesubsection{\arabic{section}.\arabic{subsection}}
\def\thesubsubsection{\arabic{subsubsection}}
\def\appendix{\setcounter{section}{0}
 \def\thesection{Appendix \Alph{section}:}
 \def\theequation{\Alph{section}.\arabic{equation}}}
 
\def\YGrule{0.4}   
\def\YGbox{6.5}    
\def\SymBoxes#1#2#3#4{\newdimen\un@t \un@t#3%
\raisebox{#1}{\rule{#2\un@t}{#4}\hskip-#2\un@t
\@tempdimb\un@t \advance\@tempdimb by-#4\@tempcntb#2\relax%
\@whilenum{\@tempcntb>0}\do{
\rule{#4}{\un@t}\hskip\@tempdimb \advance\@tempcntb by\m@ne}%
\hskip-#2\un@t \rule[\un@t]{#2\un@t}{#4}%
\rule[\un@t]{#4}{#4}\hskip-#4
\rule{#4}{\un@t}}\hskip-#4}                
\def\Young{\@ifnextchar[{\@Young}{\@Young[0]}}
\def\@Young[#1]#2{\newdimen\YG@unit \YG@unit\YGbox pt%
\newdimen\h@ight \h@ight#1\YG@unit \@tempcnta-1\relax
\@tfor\c@ount:=#2\do{\advance\@tempcnta by\@ne}
\@tempdima\@tempcnta\YG@unit%
\advance\h@ight by\@tempdima\relax     
\@tfor\c@ount:=#2\do{\SymBoxes{\h@ight}{\c@ount}{\YG@unit}{\YGrule pt}%
\@tempdima-\c@ount\YG@unit \hskip\@tempdima%
\advance \h@ight by -\YG@unit}         
\@tempdima\YG@unit \multiply\@tempdima by\@car#2\@nil %
\hskip\@tempdima}                      
\def\YoungTab{\@ifnextchar[{\@YoungIdx}{\@YoungIdx[0]}}
\def\@YoungIdx[#1]{\@ifnextchar[{\@iYoungIdx[#1]}{\@iYoungIdx[#1][\@empty]}}
\def\@iYoungIdx[#1][#2]#3{%
\newdimen\YG@unit \YG@unit\YGbox pt\newdimen\YG@rule \YG@rule \YGrule pt
\newcount\c@ount \c@ount\z@ \newdimen\skip@wd \unitlength\@ne pt
\newdimen\h@ight \h@ight#1\YG@unit \@tempcnta\m@ne\relax
\@tfor\d@um:=#3\do{\advance\@tempcnta by\@ne}
\@tempdima\@tempcnta\YG@unit%
\advance\h@ight by\@tempdima\relax
\@tfor\@idxlist:=#3\do{
\@tempcnta\z@\hskip.5\YG@rule\relax 
\@for\@idx:=\@idxlist\do{
\raisebox{\h@ight}{\makebox(\YGbox,\YGbox){#2$\@idx$}}
\advance\@tempcnta by\@ne}\hskip-.5\YG@rule%
\@tempdima-\@tempcnta\YG@unit \hskip\@tempdima%
\ifnum\c@ount=\z@ \skip@wd-\@tempdima\fi \relax
\SymBoxes{\h@ight}{\@tempcnta}{\YG@unit}{\YG@rule}%
\hskip\@tempdima \advance\h@ight by -\YG@unit
\advance\c@ount by\@ne}
\hskip\skip@wd}                      

\begin{document}

\newcommand{\beq}{\begin{equation}}
\newcommand{\eeq}{\end{equation}}
\newcommand{\bea}{\begin{eqnarray}}
\newcommand{\eea}{\end{eqnarray}}
\newcommand{\beas}{\begin{eqnarray*}}
\newcommand{\eeas}{\end{eqnarray*}}
\newcommand{\defi}{\stackrel{\rm def}{=}}
\newcommand{\non}{\nonumber}
\newcommand{\bquo}{\begin{quote}}
\newcommand{\enqu}{\end{quote}}
\def\de{\partial}
\def\Tr{ \hbox{\rm Tr}}
\def\const{\hbox {\rm const.}}
\def\o{\over}
\def\im{\hbox{\rm Im}}
\def\re{\hbox{\rm Re}}
\def\bra{\langle}\def\ket{\rangle}
\def\Arg{\hbox {\rm Arg}}
\def\Re{\hbox {\rm Re}}
\def\Im{\hbox {\rm Im}}
\def\diag{\hbox{\rm diag}}
\def\longvert{{\rule[-2mm]{0.1mm}{7mm}}\,}

\bigskip


\title{Non-Abelian Monopoles, Vortices and Confinement}

\author{Kenichi  Konishi}

\address{Dipartimento di Fisica   ``E. Fermi"  -- Universit\`a di Pisa\\
Istituto Nazionale di Fisica Nucleare -- Sezione di Pisa \\
Via Buonarroti, 2,   Ed. C, 56127  Pisa, Italy  
\\
{\it   konishi@df.unipi.it} }
\maketitle

\abstracts{Three closely related issues will be discussed. 
Magnetic quarks having  non-Abelian charges  have been found  recently to appear as  the dominant infrared degrees of
freedom   in some vacua of   softly broken ${\mathcal N}=2$  supersymmetric QCD   with $SU(n_c)$
gauge group.  Their condensation upon  ${\mathcal N}=1$ perturbation causes  confinement and dynamical symmetry
breaking.   We argue  that these magnetic quarks  can be naturally related to   the semiclassical  non-Abelian
monopoles of the type  first discussed by Goddard, Nuyts, Olive and E. Weinberg.   We discuss also general properties of  non-Abelian 
vortices and discuss their relevance to the confinement in QCD.  Finally,    
calculation by Douglas and Shenker of  the tension ratios  for vortices of different 
$N$-alities  in the softly broken ${\mathcal N}=2$  supersymmetric $SU(N)$  Yang-Mills theory,   is carried to the second order in 
 the   adjoint multiplet  mass.   
A correction to the ratios violating the   sine  formula  is  found,  showing that the latter is not   a 
universal  quantity. }

 




\section {Confining vacua of softly broken ${\mathcal N}=2$ supersymmetric QCD}   

Recently   detailed properties  of confining vacua  have been studied in a class of softly broken ${\mathcal N}=2$ supersymmetric
gauge theories.       Confining vacua in
$SU(n_c)$,
$USp(2n_c)$  or 
$SO(n_c)$  gauge theories with softly broken ${\mathcal N}=2$  supersymmetry,     with various  number of flavors   $n_f  <  2 n_c, \,\,  2 n_c+2,\,\,
n_c -2$, respectively,  have been found \cite{CKM,ArPlSei}   to fall into (roughly speaking) the following three types  (see Table \ref{tabsun}  for
the phases in
$SU(n_c)$ theories):

\begin{table}[t]
\tbl{Phases of $SU(n_c)$ gauge theory with $n_f$ flavors, taken from  
[1]. $ \tilde n_c \equiv n_f-n_c$.}
{\footnotesize  
{\begin{tabular}{|c|c|c|c|c|}
\hline
label ($r$)    &   Deg.Freed.      &  Eff. Gauge  Group
&   Phase    &   Global Symmetry     \\
\hline
$0$    &   monopoles   &   $U(1)^{n_c-1} $               &   Conf. 
   &      $U(n_f) $            \\ \hline
$ 1$           &  monopoles         & $U(1)^{n_c-1} $        &
Conf.      &     $U(n_f-1) \times U(1) $        \\ \hline
$ < [{n_f \over  2}] $  &  dual quarks        &    $SU(r)
\times U(1)^{n_c-r}   $  &    Conf. 
&          $U(n_f-r) \times U(r) $
\\ \hline
$ {n_f / 2}  $       &   rel.  nonloc.     &    -    &   SCFT
&          $U({n_f \o  2} ) \times U({n_f \o  2}) $
\\ \hline
${\tilde n}_c$     &  dual quarks     &
$ SU({\tilde n}_c) \times  U(1)^{n_c -  {\tilde n}_c } $                &
Free Mag  
&      $U(n_f) $         \\ \hline
\end{tabular}}}
\label{tabsun}
\end{table}

In some of the vacua  (the $r=0$  or $r=1$  vacua of $SU(n_c)$ theories;  also confining vacua of all flavorless cases \cite{SW1,SW2,curves}), 
the gauge group of the low-energy dual theory is  the maximal    Abelian subgroup  $U(1)^R$, where  $R$ is the rank of the original gauge group;  
confinement  is described by 't Hooft-Mandelstam mechanism \cite{TM};  

 In the general  $r$ vacua  ($2 \le r  < {n_f \o 2 } $)  of the  $SU(n_c)$ theory,   the effective low-energy theory is a   non-Abelian
$SU(r) \times  U(1)^{n-r}$  gauge  theory;   massless magnetic monopoles in the fundamental representation of  dual $SU(r) $  gauge group
appear as the low-energy degrees of freedom.  Their condensation, together that of Abelian monopoles of the  $U(1)^{n-r-1}$  factors, describes
the confinement as a generalized dual Meissner effect.  The vacua in the same universality classes  appear in  $USp(2n_c)$ and  $SO(n_c)$ theories 
with nonzero bare quark masses;

 In the $r= { n_f \o 2}$   vacua of  $SU(n_c)$ theory,  as well as in   {\it all}  of confining vacua of $USp(2n_c)$ and  $SO(n_c)$ 
theories with  massless flavor \footnote{ There are  exceptions to this rule  for small values of $n_f$ and $n_c$, 
e.g., 
 $USp(2)= SU(2)$ case.  See the
footnote 18  of 
\cite{CKM}.},  the low-energy degrees of freedom involve relatively non-local objects: the low-energy theory is a deformed superconformal
theory,
i.e., near an infrared fixed-point.

\section {  Non-Abelian Monoples }  

     We argue first   that the  ``dual quarks"   appearing in the $r$-vacua  of the softly broken ${\mathcal N}=2$ 
$SU(n_c)$ theories    can  naturally be identified with   the   non-Abelian  magnetic
monopoles of the type first discussed by Goddard,  Nuyts and Olive \cite{GNO} and studied further by E. Weinberg \cite{EW}.   Our argument is
based on the simple observations as regards to  their charges,  flavor quantum numbers,  and some general properties of electromagnetic
duality \cite{BK}.

\subsection    { \it Charges  of non-Abelian monopoles } 
Consider \cite{GNO} a broken gauge theory,
$$   G   \,\,\,{\stackrel {\bra \phi \ket    \ne 0} {\Longrightarrow}}     \,\,\, H  $$ 
where  the unbroken group  $H$  is  in general  non-Abelian.   In order to have a finite mass,  the scalar  fields 
must behave asymptotically as  
\beq    {\mathcal D} \phi    \,\,\,{\stackrel {r \to   \infty  } {\longrightarrow}}   \,\,\,0       \quad    \Rightarrow   \quad 
\phi \sim   U \cdot  \bra \phi \ket  \cdot U^{-1},
~~~   A_i^a  \sim  U \cdot {\de_i  }  U^{\dagger}   \to     
\epsilon_{aij}  { r_j 
\o     r^3}   G(r),      \eeq
with  ${\mathcal  D} G =0$,   representing nontrivial elements of   $  \Pi_2(G/H)  = \Pi_1(H)$.
The  function   $G(r)$ can be chosen as
\beq    G(r)  =      \beta_i   T_i,   \qquad   T_i  \in  {\hbox {\rm  Cartan Subalgebra of}} \,\,  H.  
\eeq
Topological quantization leads to the result that   the  ``charges"  $ \beta_i $   take values which are  
weight vectors of the group   ${\tilde H}  =    {\hbox {\rm  dual   of}} \,\, H. $
The dual of  a group (whose  roots vectors are $\alpha$'s)   is by definitioin has  the root vectors  which span dual lattice,
i.e.,    ${\tilde \alpha }= \alpha /  \alpha^2.$
Examples of pairs of the duals  are given in the Table \ref{tabtheta}
   
\begin{table}[t] 
\tbl{Some examples of dual pairs of groups}  
{\begin{tabular}{c  c   c}
\hline  
$SU(N)/Z_N       $        &   $\Leftrightarrow$                 &    $SU(N)     $          \\
  $ SO(2N)  $     &   $\Leftrightarrow$    &   $SO(2N) $       \\  
  $ SO(2N+1)  $     &   $\Leftrightarrow$     &   $USp(2N) $       \\ \hline
\end{tabular}}
\label{tabtheta}
\end{table}

As an example, consider  an  $SU(3) $ theory  broken as 
\beq     SU(3) {\stackrel {\bra \phi \ket } {\longrightarrow}}     SU(2) \times  U(1), \qquad   \bra \phi\ket = 
 \pmatrix{  v & 0& 0  \cr  0 & v & 0 \cr  0&0& -2v  }. \qquad \qquad     (\%)  
\eeq
Take a subgroup  $SU_U(2)  \subset SU(3) $
{\footnotesize    \beq    t^4= { 1\o 2}   \pmatrix{  0 & 0& 1  \cr  0 & 0 & 0 \cr  1  &0& 0   }; \quad  
t^5= { 1\o  2}   \pmatrix{  0 & 0& -i  \cr  0 & 0  & 0 \cr i &0& 0    }; \quad  
{ t^3 +  \sqrt3 t^8 \o 2}  = { 1\o 2}   \pmatrix{  1  & 0  & 0  \cr  0 & 0 & 0 \cr  0&0& -1   }; \quad  
\eeq     }    
then   
\beq      SU_U(2) {\stackrel {\bra \phi \ket } {\longrightarrow}}     U_U(1).    \qquad \qquad  \qquad  \qquad (*) \eeq
Embedding the  't Hooft-Polyakov  monopole  solution    $ \phi(r), A(r)$   for ($*$)  \cite{TP}     one gets a $SU(3)$  solution  (Sol. 1) : 
\beq \phi=
   \left( \begin{array}{ccc}
     -\frac{1}{2}v&0&0\\
     0&v&0\\
     0&0&-\frac{1}{2}v\\
   \end{array} \right)
   +
   \frac{3}{2} v \Big( t_4,t_5,\frac{t_3} {2} + \frac{\sqrt{3} t_8}{2} \Big)
   \cdot \hat{r} \phi(r), \eeq
\beq  \vec{A}=  \Big( t_4,t_5,\frac{t_3} {2} + \frac{\sqrt{3} t_8}{2} \Big)
   \wedge \hat{r} A(r). \eeq
Together with another solution  (Sol.2)  with      $SU_V(2)  \subset SU(3) $
{\footnotesize $$   t^6= { 1\o 2}   \pmatrix{  0 & 0&  0  \cr  0 & 0 & 1 \cr  0  & 1 & 0   }; \quad  
t^7= { 1\o 2}   \pmatrix{  0 & 0& 0  \cr  0 & 0  & -i  \cr 0 & i & 0    }; \quad  
{ -  t^3 +  \sqrt3 t^8 \o 2}  = { 1\o 2}   \pmatrix{  0  & 0  & 0  \cr  0 & 1 & 0 \cr  0&0& -1   }; \quad  
$$  } 
they yield a degenerate doublet of monopoles with charges   
\begin{center}
{\begin{tabular}{c  c   c}
 \\
  monopoles    &    ${\tilde  {SU}  (2)  } $           &    ${\tilde U(1) }     $          \\    \hline 
  ${\tilde q}  $        &     $ {\underline 2 }$     &      $1$       \\  \hline 
\end{tabular}}
\vskip .5cm
\end{center}

This construction can be generalized   to cases with gauge symmetry breaking 
\beq    SU(n) {\stackrel {\bra \phi \ket } {\longrightarrow}}     SU(r) \times  U^{n-r}(1), \qquad   \bra \phi\ket = 
 \pmatrix{  v_1  {\bf 1}_{r\times r}   & {\bf 0 } &  \ldots &   {\bf 0}   \cr  {\bf 0 }  & v_2   & 0   & \ldots   \cr  {\bf 0}  &0& \ddots 
&  
\ldots
\cr {\bf 0}    & 0   &  \ldots  & v_{n-r+1}   }.   
\eeq
By considering  various   $SU_i(2)$  subgroups  ($ i=1,2,\ldots, r$)   living in $[  i , r+1  ]    $  subspace  we find \\
(i) a  degenerate  $r$-plet of stable monopoles  ($ q$),  gauge (Weyl-)   transformed to each other by $SU(r) \subset SU(n)$; \\  
(ii) Abelian monopoles  ($e_i$),   ($i=1,2,\ldots,  n-r-1$)     of     $U^{n-r}(1)$  (non degenerate). 

The  charges of these stable monopoles are identical to those found in the  $\, r\,$-vacua of the softly broken  $ {\mathcal  N}=2$ 
SQCD  (Table.\ref{tabnonb})!    In particular, as will be shown in the next subsection these non-Abelian monopoles can acquire flavor quantum
numbers  through the (generalized)   Jackiw-Rebbi mechanism \cite{JR}.

\subsection {\it     Fermion  Zero modes   in   non-Abelian  monopole Background }

We now couple fermions in the fundamental representation of the gauge group.  To be concrete consider the case of a $SU(3)$ theory. 
  The fundamental multiplet, 
\beq \psi_L = \psi_{L(2)} \oplus \psi_{L(0)},  \qquad   \psi_R = \psi_{R(2)} \oplus \psi_{R(0)} \eeq
satisfies  the   Dirac equation  
$  \gamma_i {\mathcal D}_i        \psi   =0.    $   More explicitly, 
\[ -\vec{\sigma} \cdot \vec{p} \psi_{L(2)} - e \vec{\sigma} \cdot
   (\vec{t} \wedge \hat{r}) A(r) \psi_{L(2)} - \frac{1} {2} v
   \psi_{R(2)} +  3  v  \vec{t} \cdot \hat{r}
   \psi_{R(2)} \phi(r) = 0,  \]
\[ -\vec{\sigma} \cdot \vec{p} \psi_{L(0)} +  v \psi_{R(0)} = 0,  \]
\[ \vec{\sigma} \cdot \vec{p} \psi_{R(2)} + e \vec{\sigma} \cdot
   (\vec{t} \wedge \hat{r}) A(r) \psi_{R(2)} - \frac{1} {2} v
   \psi_{L(2)} + \frac{3 v}{2}  \vec{t} \cdot \hat{r}
   \psi_{L(2)} \phi(r) = 0,  \]
\beq  \vec{\sigma} \cdot \vec{p} \psi_{R(0)} +  v \psi_{L(0)} = 0.   \label{dirac}  \eeq   
Through the Yukawa coupling, the fermion acquired a mass,    $m  =   { v \o 2  } $. 
Generalizing the Jackiw-Rebbi analysis to the  massive fermions,  it can be shown that  a normalizable zero mode exists  if   $ 3  v  >    v   $ 
 which is obviously satisfied.   Each fermion gets one zero mode; quantum mechanically, the monopoles become flavor multiplets. 

An analogous construction in the case of the breaking
$ SU(n_c) \to    SU(r) \times   U(1)^{n_c -r}, $
the above condition is replaced by 
\beq   \left|  { v_0  -   v_{r+1}  \o 2}    \right| >    \left|  { v_0   +     v_{r+1}  \o 2}    \right|.    \eeq
Note that for the breaking   $SU(n) \to  SU(n-1)\times U(1)$ such a condition is always satisfied; 
otherwise, only the monopoles with  VEVS satisfying   the above condition will give rise to fermion zero modes.

This  mechanism  ``explains" the low-energy degrees of freedom in the "$r$" vacua of 
softly broken  $N=2$    SQCD, with    $G=SU(n_c)$,  with $n_f$   quarks:

\begin{table}[t]
\tbl{The effective degrees of freedom and their quantum 
numbers at a confining  $r$-vacua [2,\,1]. }   
{\begin{tabular}{|c|c|c|c|c|c|}
\hline
&   $SU(r)  $     &     $U(1)_0$    &      $ U(1)_1$   
&     $\ldots $      &   $U(1)_{n_c-r-1}$      \\
\hline\hline
$n_f \times  q$     &    ${\bf r} $    &     $1$                 &     $0$
&      $\ldots$      &     $0$                  \\ \hline
$e_1$                 & ${\bf 1}  $       &    0                    &
1      & \ldots             &  $0$              \\ \hline
$\vdots $  &    $\vdots   $         &   $\vdots   $        &    $\vdots   $
&             $\ddots $     &     $\vdots   $      
\\ \hline
$e_{n_c-r-1} $    &  ${\bf 1}  $    & 0                     & 0
&      $ \ldots  $            & 1                \\ \hline
\end{tabular}}
\label{tabnonb}
\end{table}

\subsection {  \it  Duality} 

It is also  significant that, in the softly broken ${\mathcal N}=2$  $SU(n_c)$ theory,    the $r$ vacua  with a  magnetic $SU(r)$  gauge group 
occur only for   $r  \le   {n_f \o 2}$.   
This       is a manifestation   of the   fact  that  the   quantum  behavior of
non-Abelian monopoles depends crucially on the massless matter fermion degrees of freedom in the fundamental theory.   
Indeed, the magnetic $SU(r)\times U(1)^{n_c-r}$   theory  with these matter multiplets is  infrared-free (i.e., non asymptotic free).  This is the
correct behavior as it should be  dual to the original asymptotic free $SU(n_c)$  gauge theory.  
Note that  the gauge coupling constant  evolution, which appears as  due to the perturbative loops of magnetic monopoles, 
is actually the result of, and equivalent to,    the infinite  sum of instanton contributions in  the original  $SU(n_c)$   theory.    

This is perfectly analogous to  the observation \cite{DPK}   about  how the old paradox related to the   Dirac quantization condition  and
renormali\-z\-ation  group
\cite{Zum} :  
\beq     g_e (\mu)  \cdot g_m(\mu) = 2 \pi n, \qquad \forall \mu, 
\eeq   
  is solved within  the 
$SU(2)$  Seiberg-Witten theory. 

Note that this also     explains why in the  pure   ${\mathcal N}=2$  $SU(n_c)$  theory or on a generic point of the Coulomb branch of
the  ${\mathcal N}=2$ SQCD,  the
low-energy effective theory is an  Abelian  gauge  theory \cite{SW1}-\cite{curves}.   Massless fermion flavors  
are   needed in order for   non-Abelian monopoles  to  get dressed, via a generalized Jackiw-Rebbi mechanism discussed above
with  a non trivial  
$SU(n_f)$   flavor quantum numbers  and, as a result,  to render  the dual gauge   interactions  infrared-free. 
When this is not possible,   non-Abelian monopoles are 
strongly coupled and do not manifest themselves as identifiable low-energy degrees of freedom.   

In this respect,   it is very interesting that the boundary   case $r= {n_f \o 2}$   also occurs 
(confining vacua of type (iii) discussed in Introduction)  within the class of supersymmetric theories considered in \cite{CKM}.   In these
vacua,  non-Abelian  monopoles and dyons are strongly coupled, but  still  describe   the low-energy dynamics,  albeit via  non-local effective
interactions.  
  
Non-Abelian monopoles are  actually  quite elusive objects.  Though their presence may be detected    in a semi-classical approximation, 
their true nature   depends on the long distance physics.    If the ``unbroken" gauge group  is dynamically  broken
further in the infrared  such multiplets of  states  simply represent an  approximately degenerate  set of magnetic  monopoles.  Only if  
there is no further dynamical breaking do  the non-Abelian monopoles transforming  as nontrivial multiplets of the unbroken,  dual gauge
group,  appear in the theory.

 There are strong indications that this  occurs   in the $r$-vacua (with an effective $SU(r)\times
U(1)^{n_c-r}$ gauge symmetry) of the softly broken ${\mathcal N}=2,$    
$SU(n_c)$ supersymmetric QCD \cite{CKM}.  If our
idea  is correct,  this is perhaps the first
 physical  system known  in which Goddard-Nuyts-Olive-Weinberg  monopoles manifest themselves as  infrared degrees of freedom,
playing an essential dynamical role. 
 For more about the subtle nature  of nonAbelian monopoles,  see \cite{BK}.

\section{    Non-Abelian   Vortices}  

A closely related issue is that  of non-Abelian vortices \cite{NO}-\cite{Kneipp}.  If confinement is to be described as a sort of non-Abelian 
dual Meissner effect,  the magnetic monopoles of the type discussed above condense and break the dual gauge group. As a result, 
the system develops vortex configurations which serve as confining strings.

\subsection   {\it  General Characterization}

This time we consider a gauge theory  in which the gauge group $G$ is spontaneously broken by the Higgs mechanism as 
\beq   G \Longrightarrow     {\mathcal C} 
\eeq
with $  {\mathcal C} $   a {\it discrete}  center of the group.   The general properties of the vortex, which represents a nontrivial elements of the 
fundamental group,
\beq    \Pi_1(G/C)  = C, 
\eeq
   are  
independent of the detailed form of the scalar potential or of the number of  the Higgs fields present.
 Asymptotic form of the fields are: 
$$   A_i  \sim   { i \o  g}    \, U(\phi) \de_i  U^{\dagger}(\phi);    
\quad     \phi_A \sim   U \phi_{A}^{(0)}  U^{\dagger},   \qquad  U(\phi) = \exp{i \sum_j^r 
\beta_j T_j
\phi}
$$
where  
   $T_i$'s
can be taken in the     Cartan subalgebra of $G$: then  
$$   A_{\phi}  \sim   { 1 \o  g  \, r  }  \sum_j^r  \beta_j T_j 
$$
The vortex flux   
$$  \oint  dx_i A_i    =  { 2 \pi  \o g   }   \sum_j^r  \beta_j T_j,   
$$
is characterized by  the ``charges"  $\beta$.    The quantization condition
\beq   U(2 \pi)  \in    {\mathcal C}.     \label{qcondi1}  \eeq
leads to the 
result that  $\beta_j$'s     are      weight vectors of   $ {\tilde G} $   (dual of   $G$). 
 $\beta_j$'s  are actually 
defined modulo  Weyl  transformations
$\beta_i$'s: 
$$   
  \beta^{\prime}=  {\bf \beta } - { 2 {\bf \alpha}  ( {\bf \beta  }\cdot {\bf \alpha}) \o  ( {\bf \alpha} \cdot  {\bf \alpha})},
\label{weyltr} $$ 
where  $ \alpha$ is a root vector of $G$.   

\subsection{ \it  $SU(N)/Z_N$}

The simplest system with non-Abelian vortices is  {  $\mathbf  {SO(3)=SU(2)/Z_2}$   } 
 broken to ${\mathbf  Z}_2$.   It has 
\begin{itemize} 
\item      unique   $Z_2$   vortex ( the source charge additive {\it mod 2} ); 
\item   ``  flux " 
$$       \int_{S} { F}_{ij}^3 \, d\sigma_{ij} =   \oint dx_i    { A}_{i}^3  =   { 2 \pi n \o g}. 
\label{winding}  $$ 
which  is conserved but  not gauge invariant.      $n=2$ ``vortex"  can be gauge-transformed away. 
\end{itemize}

A more interesting system is   {  $\mathbf {SU(3)/Z_3 }$ }, i.e.,  $SU(3)$  theories with all fields in 
adjoint representation. The Cartan subalgebra can be taken to be 
  {\footnotesize 
$$   T_3= { 1 \o 2 \sqrt3} \pmatrix {1 & 0 & 0\cr 0 & -1 & 0\cr
0 & 0& 0};\qquad T_8= { 1 \o 6}  \pmatrix {1 & 0 & 0\cr 0 & 1 & 0\cr
0 & 0& -2}.$$  }  
The quantization condition  $  U(2\pi) \in Z_3$    leads to the equations 
\beq  { \beta_3 \o 2\sqrt3 } +  { \beta_8 \o  6} =  - { n_1 \o 3}, \quad   -{ \beta_3 \o 2\sqrt3 } +  { \beta_8 \o  6} =  - { n_2 \o 3},
 \quad
   - { 1\o 3}   \beta_8=  - { n_3 \o 3},   \label{quantiz1}  \eeq 
to be solved with the condition, $  \sum_{i} n_i=0.$
The simplest    $ N$-ality (triality)   one   ($      n_i=    {[1 \,\,  mod \,\,    3] }$)       solutions are: 

\beq   {\bf \beta}= ( - \sqrt 3, 1 ), \,\,  (  \sqrt 3,  1 ), \,\, {\hbox{\rm or}}    \,\,   (  0, - 2 ) \,\, = 2 N   { {\bf w }}
\label{minsol} \eeq
${\bf w }$   $=$    weight vector of   ${ {\bf 3}}$.    Thus the sources of the minimum vortex carry the quantum number of 
the quarks.  The dual of the theory we are studying,  {  ${SU(3)/Z_3 }$ }, is indeed   {  ${SU(3)} $}!   

By adding four of  (\ref{minsol})  $\to $   ${ {\bf 6}}^{*}$   ($ \Young [-0.5] { 2 2 }$ ),      etc., and one could construct 
an infinite number of triality-one solutions.  
However only the vortex with the  lowest tension  is stable.

\begin{figure}[h]
\begin{center}
\epsfig{file=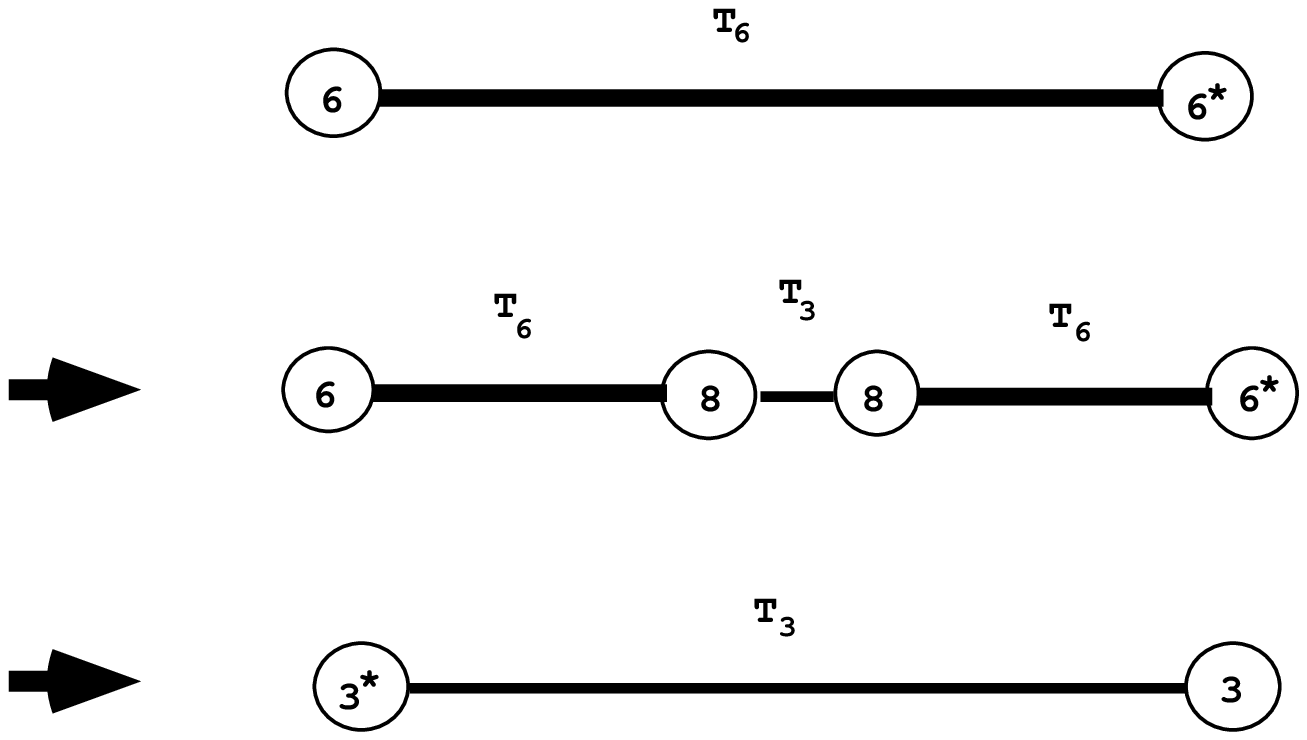,  width=8cm}                
\end{center}
\caption{ }    
\label{vortex}    
\end{figure}

 ${ N}$-ality (triality)-two solutions are found by adding vectorially the minumum solutions above.
   The source of these vortices  correspond to the irreducible representations     has charges 
$$   {\bf \beta}= (  -\sqrt 3, -1 ), \,\,  (  \sqrt 3, -1 ), \,\, {\hbox{\rm or}}    \,\,   (  0, 2 ), \,\, 
$$
$=$  weight vectors of    ${\underline  3^{*}}$      ($\Young [-0.5]  { 1 1 } $),      or 
   $$   {\bf \beta}= (  -2 \sqrt 3, 2 ), \,\,  (  2\sqrt 3, 2 ), \,\,   (  0, 2 ).  \,\,    ( - \sqrt 3,  -1 ), \,\,   (  \sqrt 3, -1 ), \,\, 
{\hbox{\rm or}}   
\,\,  (  0, -4 ), \,\,
$$ 
$=$    ${\underline  6}$   ($ \Young {2 }$).
  Quantum mechanically,   however,  
the vortex with the higher tension   (probably  ${\underline  6}$)     decays   through
the  gauge  boson pair productions (Fig. \ref{vortex}).  
Somewhat similar problem of decay of metastable vortices  was recently discussed by Shifman and Yung \cite{SY}.

\begin{figure}[h]
\begin{center}
\epsfig{file=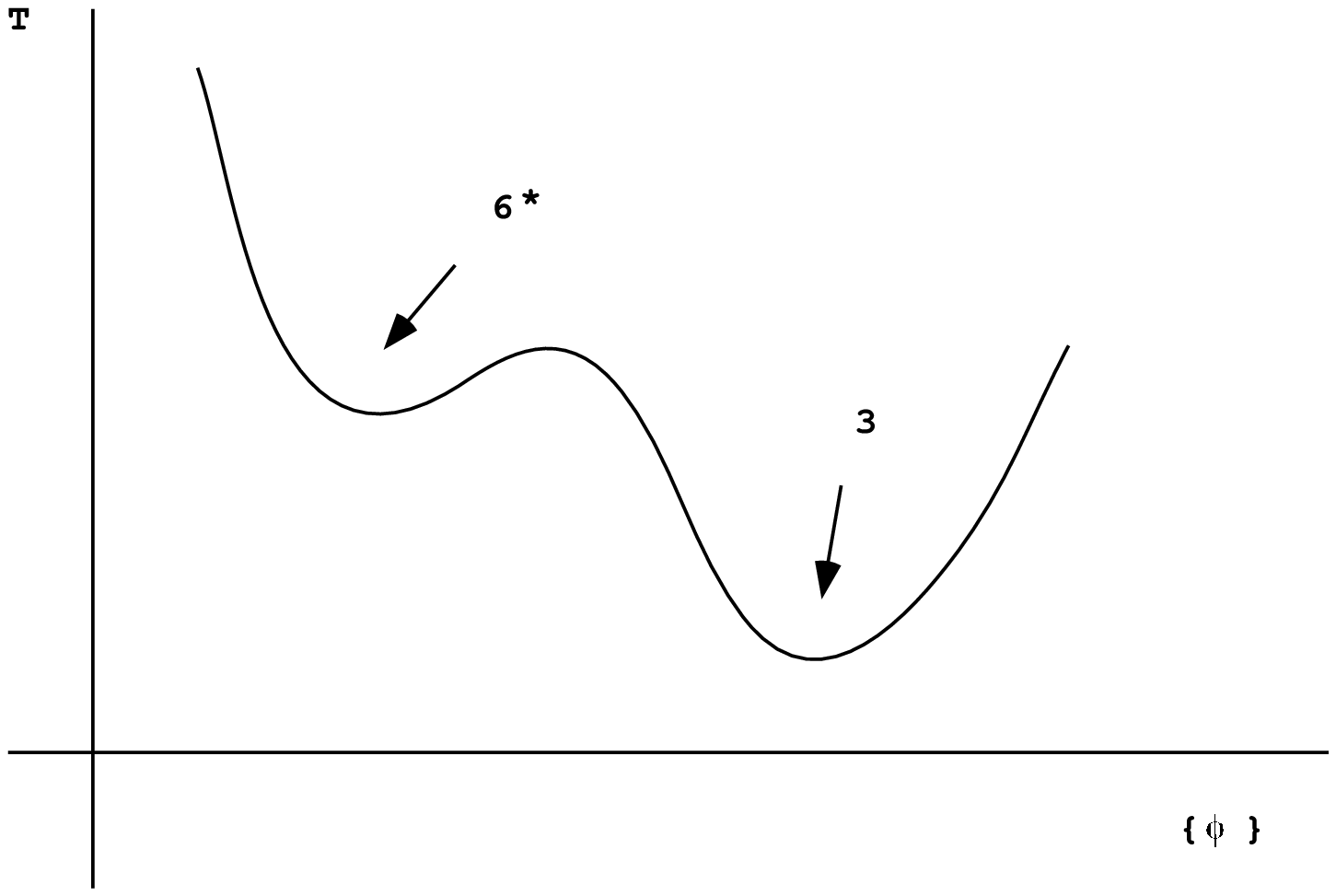,  width=9cm}                
\end{center}
\caption{ }       
\label{}    
\end{figure}

Since   ${\underline  3^{*}}$  vortex and  ${\underline  3}$ vortex  are equivalent,  there is actually a unique  stable vortex with 
minimum $\mathbb Z_3$  charge in  the   $SU(3)$ gauge theory.

The discussion can be generalized natually to {    ${\mathbf {SU(N)/Z_N}}$    }   theory in Higgs phase. 
One finds  $N$ degenerate solutions  of    {$N$-ality one  }
$$    {\bf \beta}_j =     2 N  {\bf w}_j, \qquad j=1, 2,3,\ldots,  N      \label{betasol}   $$
\beq
  {\bf w}_j = {\hbox{\rm   the weight vectors of the  }}    \, {\underline N}. 
\eeq
There are also solutions  representing   vortices of higher $N$-alities.     At  the $N$-ality two, for
instance,   the  solutions for $\beta$ have   the form, 
\beq      2 N ({ {\bf  w}}_i  + { {\bf  w}}_j), \qquad    i,j = 1,2, \ldots, N.   
\eeq 
They fall into  two gauge inequivalent sets of  vortices:   their sources would    carry  the quantum numbers of the two
irreducible representations,   
\beq      \Young {2 } \,,  \quad  \Young [-1] { 1 1 }\, , 
\eeq   
symmetric and antisymmetric in color, respectively.

Solutions of   $N$-ality   $k$   can be analogously be constructed  by taking as $\beta$ the vector sum of  arbitrary $k$ 
minimum solutions,   Eq.(\ref{betasol}).    These vortices can be grouped  into gauge invariant subsets, each of which has a source 
  carrying      quantum numbers 
of an      irreducible  representations of $SU(N)$ group,  
\beq    \overbrace{\Young {3 } \ldots  \Young {2 }}^k, \qquad   \overbrace{\Young [-1]{2  1} \ldots   \Young {2 }}^{k-1},\qquad 
\ldots,   \qquad  \Young[-4]{1 1 1 1 1} \, , 
\label{allthe} \eeq   
all having $k$ boxes.  

The vortices of $N$-ality, $1\le k \le   N-1,  $   cannot be  unwound   by a gauge transformation.
    Nevertheless,   this does not mean that each of  the   vortices (\ref{allthe})    is  stable against decay.    A  vortex of a given $N$-ality can decay through the 
pair production  of gauge bosons   into one  of the same  $\mathbb Z_N $  quantum number but 
 with a lower   tension, via processes   
similar to the one in the $SU(3)$  example of   Fig. \ref{vortex}.         It is possible that   the tension 
is  smallest in the case of  the antisymmetric
representation
$\underline { N   \choose  k}.$      If it is so,  the solution for the vortex charge $\beta$     at $N$-ality
$k$ is truely  a unique gauge-invariant set
\beq   2 N \,   \{  { {\bf  w}}_{i_1}   + { {\bf  w}}_{i_2}  + \ldots + { {\bf  w}}_{i_k} \quad  mod    \quad    \alpha \},   \qquad    i_m = 1,2, \ldots,
N,   
\eeq 
where $\alpha$'s are the root vectors of the $SU(N)$ group.  
  These represent
$$     \Pi_1 (SU(N)/Z_N) = Z_N. 
$$
  
Which of these, apart from the smallest,  $N$-ality one  vortex,  is stable against decay into   a  bundle of 
vortices with smaller   $N$-alities,    is again  a  dynamical
question   (i.e., depends on  the form of the potential, values of  coupling constants,  quantum corrections, etc.).  
    One would expect   no  universal      formula for the relative tensions
among vortices of different $N$-alities,      on the general ground.  However, 
  there are  some intriguing    suggestions \cite{STR}  
that  the ratios  among the vortex tensions  for  different $\mathbb  Z_N$ charges,   found originally    in the pure ${\mathcal N}=2$
supersymmetric Yang-Mills theory  (broken softly to ${\mathcal N}=1$) \cite{DS},  
\beq      {T_k \o T_{\ell}} =      { \sin { \pi   k \o  N} \o \sin { \pi   \ell \o  N}}, 
\label{universal?}  \eeq
might be  universal.     
The  results  from lattice
calculations   with $SU(5)$ and $SU(6)$  Yang-Mills theories      
\cite{LT,Pisa}  are consistent with the sine formula.  More recent results on these ratios  \cite{KI,KA} however 
seem  to  indicate  the non-universality of these ratios.

The absence of   vortices  of $N$-ality, $N$,  can be understood since  the  charges   corresponding to an irreducible representation
with $N$ boxes  in the Young tableau,   can always be    screened by those of the dynamical fields (adjoint representation):
 the
vortex is broken by copious production of massless gluons   of the dual $SU(N)$  theory.

In an analogous fashon,   one finds that 
sources of vortices  in $ \mathbf {USp(2N)}$  theory in Higgs  phase carry the weights of the   $2^{N}$  dimensional  spinor representation of  
the dual group,
$SO(2N+1); $
sources of  vortices  in $\mathbf {SO(2N+1)}$  theory in Higgs  phase carry the weights of the   $2N$  dimensional fundamental representation
 of   $ {USp(2N)}$, etc.   For more details, see \cite{KSp}.

\subsection  {\it  Remarks}

{(i)} If  confinement   in $SU(N)$  theory 
   is a  dual   Meissner effect   with        Olive-Montonen  duality,
 $  SU(N) \leftrightarrow
SU(N)/Z_N,   $ 
then  the universal  $q-{\bar q}$  meson Regge trajectory  will be naturally explained,   in contrast to the case when the dual theory is  
$U(1)^{N-1}$; \\ 
{(ii)}   Sources of the non-Abelian vortices have  charge additive only {\it mod $N$}.  Non-Abelian vortices are  
   non BPS:  linearized approximation is not valid in general; \\
{(iii)}   Explicit construction of non-Abelian vortices  \cite{NO}-\cite{Kneipp}
has been studied by using  simple models for the adjoint scalar potential.  However a systematic study of 
non-Abelian vortices, hence of non-Abelian superconductors, are still lacking.\\ 
{(iv)}   What is the relation between vortex formation and XSB? \\
{(v)}   Can we compute the ratios   of  vortex tensions for different $N$-alities (in the $SU(N)$ case)? 

This last point brings us    to our third  issue,   related to Eq.(\ref{universal?}).

\section {  Non-Universal Corrections in the Tension Ratios in softly broken ${\mathcal N}=2$  $SU(N)$
Yang-Mills   }

Derivation  of formula  such as  Eq.(\ref{universal?})  in the standard, continuous $SU(N)$   gauge theories   still defies us.    The first
field-theoretic  result  on this issue was obtained by   Douglas and Shenker \cite{DS},    
  in the  ${\mathcal N}=2$ supersymmetric  $SU(N)$  pure Yang Mills theory, with supersymmetry   softly broken to  ${\mathcal N}=1$  by a small  
adjoint scalar multiplet mass  $m$. 
They  found  Eq.(\ref{universal?})  for the  ratios
of   the   tensions   of    abelian    (Abrikosov-Nielsen-Olesen) \cite{ABR} vortices corresponding to   different  $U(1)$ factors of
the low-energy  effective (magnetic)  
$U(1)^{N-1}$   theory.    

 The  $n$-th  color component of the quark has  charges 
\beq       \delta_{n, k} - \delta_{n, k +1}, \qquad   (k=1,2,\ldots, N-1; \, n=1,2,\ldots, N) 
\eeq
 with   respect to the various  electric $U_k(1)$  gauge groups.       The source   of the 
$k$-th   ANO string thus  corresponds to the $N$-ality $k$   multiquark state,
$|k \ket  = |q_1 q_2,\ldots q_k \ket$, allowing a re-interpretation of Eq.(\ref{universal?})  as referring to the
ratio of the tension for different $N$-ality  confining strings \cite{STRASS}.   
  
   However,    physics of  the softly broken  ${\mathcal N}=2$  $SU(N)$  pure Yang-Mills theory is   quite different from
what  is expected in QCD.     Dynamical   $SU(N) \to U(1)^{N-1}$ breaking  introduces multiple
of   meson Regge trajectories with different slopes at low masses \cite{STRASS,YUNG}, a feature which is neither  seen in Nature nor expected in
QCD.  For instance, another
$N$-ality
$k$ state   
$|k\ket ^{\prime}= |q_2 q_3,\ldots q_{k+1} \ket$ acts as  source of the  $U_{k+1}(1)$ vortex and as the  sink of the
$U_2(1)$ vortex, which together bind $ |k \ket^{\prime}$-  anti $  |{k} \ket^{\prime}$ states with   a  tension different from $T_k$. 
The  Douglas-Shenker prediction is, so to speak, a good  prediction for a wrong theory! 
Only in the limit of ${\mathcal   
N}=1$ does one expect  to find one  stable vortex for each $N$-ality,   corresponding to the conserved $Z_N$ charges \cite{STRASS}.

Within the softly broken ${\mathcal N}=2$  $SU(N)$ theory,   the two regimes can  be in principle smoothly interpolated
by varying the adjoint mass $m $ from zero to  infinity, adjusting appropriately $\Lambda$.  At small $m $  one has a good
local description of the low-energy effective  dual, magnetic  $U(1)^{N-1}$   theory.      The transition towards large 
$m$   regime involves both perturbative and  nonperturbative effects.    Perturbatively, there are higher  corrections
due to the ${\mathcal N}=1$ perturbation, $m   \, \Tr \Phi^2$.   Nonperturbatively - in the dual theory -   there are productions of massive gauge 
bosons  of  the broken $SU(N)/U(1)^{N-1}$ generators,    which mix  different $  U(1)^{N-1}$ vortices
and  eventually  lead   to the unique stable vortex  with  a given ${\mathcal N}$-ality.  

Below   is   the result on  the    perturbative corrections  to the tension
ratios Eq.(\ref{universal?}), due to the next-to-lowest contributions in $m $.  We shall find a small non-universal 
correction to the sine formula Eq.(\ref{universal?}).     Our point is     not that such a 
result is of interest in itself as a physical prediction but that it gives a strong indication for  the non-universality  
of this formula, even though  it could be an approximately a good one.

 The problem of the next-to-lowest contributions in $m $ has been already  analized  in
$SU(2)$ theory, by Vainshtein and Yung \cite{YUNG} and by Hou \cite{HOU}, although in that case  there is only one $U(1)$ factor.    When
only up to the order
$A_D$  term in the expansion
\beq m \,    \bra \Tr\,  \Phi^2 \ket  =  m \,  U (A_D)  =  m \, \Lambda^2 ( 1 -  { 2 i A_D \o \Lambda}  -  { 1\o 4}  { A_D^2 \o \Lambda^2} + \ldots  ) 
 \eeq   
is kept,   the effective low energy theory turns out to be  an   ${\mathcal N}=2$   SQED,  $ A_D$  being an ${\mathcal N}=2$ analogue  of 
the Fayet-Iliopoulos term.  As a result,   the vortex remains  BPS-saturated, and its  tension is proportional to the 
 monopole charge \cite{YUNG,HOU}.   
When  the $ A_D^2$  term is    
taken into account,  the vortex   ceases to be BPS-saturated:  the correction to the vortex tension can be calculated
perturbatively, giving rise to the results  that the vacuum behaves as a type I superconductor.

 Our aim here is  to  generalize these analyses   to $SU(N)$ theory.   In fact,  
Douglas-Shenker result Eq.(\ref{universal?})  in $SU(N)$ theory  was obtained   in the BPS approximation, by keeping only  the  linear
terms  in $a_{Di}$    in  the expansion 
\beq   \,  U (a_{Di})   =   U_0 +  U_{0k} \, a_{Dk} +  { U_{0 mn}\o 2}   \, a_{Dm}\,  a_{Dn} + \ldots,  
\qquad     U_{0k} = - 4 \,i  \,\Lambda \sin{ \pi k \o N}.  
\eeq  
The coefficients $U_{0k}$ were computed by Douglas-Shenker \cite{DS}.
Our first task is then to compute the coefficients of the second term  $U_{0 mn}$.  In principle it is a straightforward matter, 
as one must  simply  invert  the Seiberg-Witten formula:\footnote {We follow  the notation of \cite{DS},  
with   $y^2=  P(x)^2  - \Lambda^2$;   $  P(x)=  { 1\o 2}   \prod_{i=1}^N (x - \phi_i)$   } 
\beq 
a_{Dm} = \oint_{\alpha_m} \lambda, \qquad a_{m} = \oint_{\beta_m} \lambda,\qquad 
\lambda =  { 1 \o 2 \pi i} { x \o y}  { \de P (x) \o \de x} dx, 
\eeq
 which is explicitly known, to second order.  The only trouble is that  $ a_{Dm} $ and $a_{m} $ ($m=1,2,\ldots, N-1$)    are given simply  in
terms of $N$  dependent vacuum parameters
$\phi_i, $   $\sum_{i=1}^N \phi_i =0$.    By denoting the formal derivatives with respect to $\phi_i$ as  ${ \delta \o \delta \phi_i}$,  one finds 
\beq   \sum_{i=1}^N  {\delta   a_{Dm} \o \delta \phi_i } { \de \phi_i \o \de \,a_{Dn}   } =   \delta_{mn}, \qquad 
\sum_{m=1}^{N-1}   { \de \phi_i \o \de \,a_{Dm}   }  {\delta   a_{Dm} \o \delta \phi_j } =   \delta_{ij} - { 1\o N}, 
\label{rela} \eeq
which follow easily by using the constraint,  $\sum_{i=1}^N \phi_i =0$.  In terms of 
$ B_{mi}\equiv  -i  { \delta  a_{D m} \o \delta \phi_i}$,   $  A_{mi}\equiv  -i  { \delta  a_{m} \o \delta \phi_i}$   which are 
explicitly given at the $N$ confining vacua in \cite{DS},    one then finds   
\beq   {\de \phi_i  \o \de \, a_{D m} } = - i B_{mi};  \qquad   \sum_{i=1}^N B_{mi} B_{ni} = \delta_{mn}; \qquad 
\sum_{m=1}^{N-1}  B_{mi} B_{mj}  = \delta_{ij} - { 1\o N}.
\eeq
The explicit values of $B_{mi}$ are (see \cite{DS}):
\begin{equation} 
 B_{mi}= \frac{1}{N} \frac{ \sin[\widehat{\theta}_m ]} {\cos [\theta_i] - 
\cos[\widehat{\theta}_m]  }; \qquad \hat{\theta}_n=\frac{\pi n}{N};  \qquad
\theta_n = \frac{\pi (n-1/2)}{N}.  \end{equation} 
The definition of $u(a_{Di})$ is the following:
\beq u(a_{Di})=\sum_i \phi_i^2. \eeq
Then the desired coefficients can be found by the following expression,
computed at  $a_{Di}=0$: 
\begin{equation} \label{grossa} U_{0mn}=\frac{\partial
^2 u}{\partial a_{Dm}
\partial a_{Dn}} = 2 \sum_k    
 \frac{\partial \phi_k}{\partial a_{Dm}} \frac{\partial \phi_k}{\partial
a_{Dn}} + 2 \phi_k \frac{\partial ^2 \phi_k}{\partial a_{Dm} \partial
a_{Dn}}.    
\end{equation} 
The first part of Eq.(\ref{grossa}) becomes:
\begin{eqnarray} 
&2 \sum_k
 \frac{\partial \phi_k}{\partial a_{Dm}} \frac{\partial \phi_k}{\partial
a_{Dn}}=- 2 \sum_k B_{km} B_{kn}= \nonumber \\
&- 2 \sum_{k,s} \frac{2}{N}\sin \left[
 \frac{\pi m s}{N} \right]
  \sin \left[
 \frac{\pi n k}{N} \right] \delta_{ks} = - 2 \delta_{mn}.
 \end{eqnarray}
The evaluation of the second term is a little tricky \cite{KA}.   The result is however simple:
\beq 2 \sum_k \phi_k \frac{\partial ^2 \phi_k}{\partial a_{Dm}
 \partial a_{Dn}} = \left( 2-\frac{1}{N} \right) \delta_{mn},
\label{calcolone} \eeq 
thus
 \beq       U_{0 mn}  =  (-   { 1\o N} )  \, 
\delta_{mn}. \label{result} \eeq

    We now use this  result to calculate the corrections to the tension ratios  (\ref{universal?})   found in the lowest order. 
The effective  Lagrangean  near one of the $N$   confining ${\mathcal N}=1$ vacua    is 
\begin{eqnarray} \label{lagry}
\lefteqn{\mathcal{L} = \sum_{i=1} ^{N-1} Im \left[
\frac{i}{e_{Di}^2} \left( \int d^4 \theta A_{Di} A_{Di}^+ + \int
d^2 \theta (W_{Di})^2 \right)      
\right]+ {} } \nonumber\\
 & & {} +Re \left[\int d^4 \theta (M_i^+ e^{V_{Di}} M_i
+\tilde{M}_i^+ e^{-V_{Di}} \tilde{M}_i)\right] + {} \nonumber\\
 & & {} + 2 Re \left[ \sqrt{2} \int d^2 \theta A_{Di} M_i \tilde{M}_i
+ m   \,  U[A_{Di}]
 \right].  
\label{lagr}  \end{eqnarray}
The coupling constant ${e_{Di}^2} $   is formally vanishing, as 
\[  \frac{4 \pi}{e^2_{Dk}}\simeq \frac{1}{2 \pi} \ln \frac{\Lambda
\sin(\widehat{\theta}_k)}{a_{Dk} N} \]    where  $\widehat{\theta}_m\equiv { \pi n \o N}  $  and  
$a_{Dk}=0$ at the minimum. Physically, the monopole loop integrals
are in fact cut off by masses caused  by the $\mathcal{N}=1$ perturbation.
The monopole  becomes massive  when  $ m \ne 0$, and   $\sqrt2 a_{Dk}$ should be replaced
by the physical monopole mass   $ (m\Lambda \sin( \widehat{
\theta}_k))^{1/2}$  which acts as the infrared cutoff for the coupling constant evolution. 
Thus 
\begin{equation} \label{accoppiamenti}
e^2_{Dm}\simeq \frac{16 \pi^2}{\ln(\frac{\Lambda \sin(
\widehat{\theta}_m)}{ m \,  N^2})}.
\label{effectivee}  \end{equation}

As $ U_{0 mn}  $ is found to be diagonal,  the description of the ANO vortices \cite{ABR,NO}  in terms of effective magnetic Abelian theory
description continues to be valid for each $U(1)$ factor.    In the linear
approximation $U(A_D) =  m \Lambda^2  +  \mu   A_D, $  where  $\mu \equiv | 4
\,  m  \,\Lambda \sin{ \pi k \o N}|$  for the $k$-th   $U(1)$ theory,  the
theory can be (for the static configurations)    effectively reduced  to an
${\mathcal N}=4$   theory in $2+1$ dimensions. In this way, Bogomolny's equations
for the BPS vortex can be easily found from the condition that the vacuum to
be supersymmetric: \begin{equation} F_{12}=\sqrt{2}\, (\sqrt{2}M^+ \tilde{M}^+
- \mu ) \qquad  (D_1 + i D_2) M = 0 \end{equation} \begin{equation}
M=\tilde{M}^+,  \qquad    A_D=0.  \end{equation} The solutions of these
equations are similar to the one considered by Nielsen and Olesen: 
\begin{equation} M=\left( \frac{\mu}{\sqrt{2}} \right) ^{1/2} e^{i n \phi} f[r
e \sqrt{\mu}], \qquad  A_{\phi}=-2 n \frac{g(r e \sqrt{\mu} )}{r}
\label{unperturb} \end{equation}
where 
\beq  f'=\frac{f}{r}(1-2g)) n \qquad 
g'=\frac{1}{2 n} r (1-f^2)  \eeq
with boundary consitions  $ f(0)=g(0)=0 $, $f(r\rightarrow
\infty)=1$, $g(r \rightarrow \infty)=+1/2 $). The tension  turns out to be independent of the coupling constant: for the minimum 
vortex 
 \beq T=\sqrt{2}
\pi \mu =  4 \sqrt{2} \pi \,  |m  \,\Lambda|  \sin{ \pi k \o N}. \label{tension} \eeq
That the absolute value of   $m$ appears in Eq.(\ref{tension}) as it should, and also  in Eq.(\ref{Result}) below, 
is not  obvious.  This can actually  be shown by an  appropriate redefinition of the field variables, used in \cite{DPK},  which
renders  all equations real.  

\begin{figure}[h]
\begin{center}
\epsfig{file=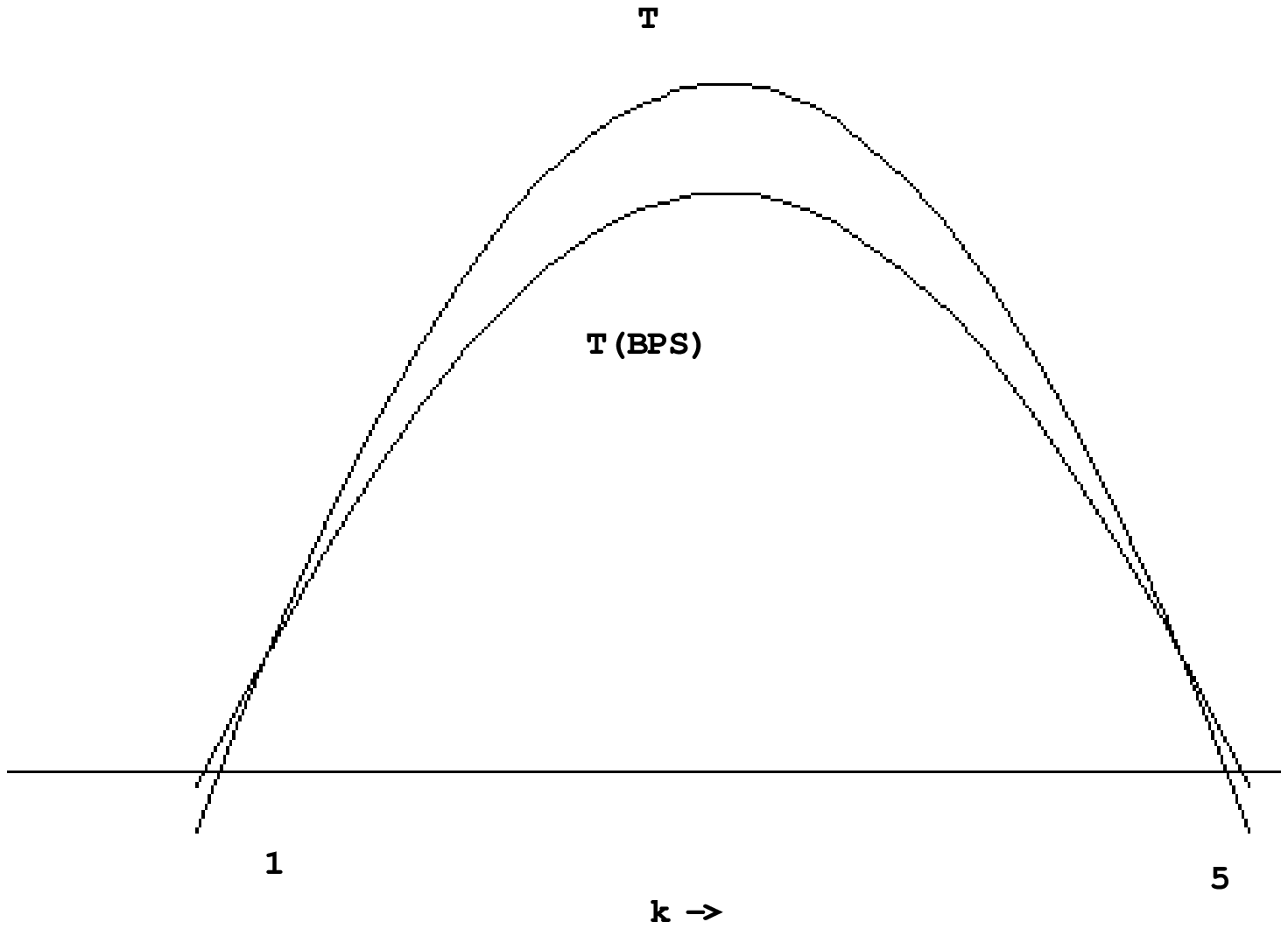,  width=6cm}                
\end{center}
\caption{ }    
\label{vortex11}    
\end{figure}

When the second order term in $U(A_D)= \mu A_D+\frac{1}{2} \eta A_D^2$,  $\eta \equiv U_{kk},$  is taken into account,  
the vortex ceases to
be BPS saturated.   The corrections to the vortex  tension  
due to $\eta$  can be taking into account by perturbation theory,  following  \cite{HOU}.  To first order, the equation for $A_{Dk}=A_D$ is 
\begin{equation}
\nabla^2 A_{D}= - 2 e^4 \eta \, (\mu -\sqrt{2} M \tilde{M}) + 2
e^2 A_{D}  (M M^+ + \tilde{M} \tilde{M}^+)
\end{equation}
where unperturbed expressions  from Eq.(\ref{unperturb}) can be used for $M$, $\tilde{M}. $  The vortex tension becomes simply
\begin{equation}
T=\int d^2 x \, [ \, (-\sqrt{2} \mu F_{12})-2 e^2 \eta A_D (\mu - \sqrt{2}
M^+ \tilde{M}^+) \, ]
\end{equation}
where the second term represents the correction.   By restoring the $k$ dependence, we finally get for the 
tension of the $k$-th   vortex, 
\begin{equation} \label{mipiace}
T_k =4 \sqrt{2}\, \pi \,  |m| \, \Lambda \sin \left(  \frac{\pi k}{N} \right)
-C \frac{   16 \pi^2   |m|^2 }{N^2 \ln \frac{\Lambda \sin(k \pi /N)} {|m|  \, 
N^2 }},  \label{Result} 
\end{equation}
where  $C=2 \sqrt{2} \pi (0.68)=6.04. $   The correction term  has a negative sign, independently of the phase of the adjoint mass. 
 Note that the relation $T_k = T_{N-k}$  continues to  hold.  
        Eq.(\ref{Result})  is valid for $m \ll \Lambda$. 
Qualitative feature of this correction is shown in Fig.\ref{vortex11},  for $N=6$.

  In the above  consideration,   we have taken into account exactly the $m^2 $ corrections in the F-term of the
effective low-energy action.  On the other hand,  the corrections to the D-terms are   subtler.   Indeed,
based on the physical consideration,   $a_D$  in the argument of the logarithm in the effective low energy coupling constant was replaced by the
monopole mass, of  $O(\sqrt {m \Lambda}).$  This amounts to the $m$ insertion to all orders in the loops.  Such a resummation is necessitated by
the infrared divergences and represents a standard procedure. Another well-known example is the chiral perturbation theory
in which quark masses appear logarithmically, e.g., in the expansion of the quark condensate.   
This explains the non-analytic dependence on
$m$ as well as on 
${ 1\o N}$  \cite{Fer}.    

  Also, there are  corrections due to nondiagonal elements  in the coupling constant matrix 
$\tau_{ij}$, which mix the different $U(1)$
factors
\cite{Edel},  neglected in Eq.(\ref{lagr}).  These nondiagonal elements  are suppressed by  $O({1 \o \log{\Lambda/m}})$ relatively to the diagonal 
ones,  apparently of the same order of suppression as the correction calculated  above. However,   these nondiagonal elements
gives rise to  corrections  to  the tension  of one order higher,  $O({1 \o
\log^2{\Lambda/m}}),$   hence is negligible to the order considered. 

We thus find a non-universal  correction to the Douglas-Shenker formula,  Eq.(\ref{universal?}).   In the process of 
transition towards fully non-Abelian superconductivity  at large $m$   nonperturbative effects such as the $W$ boson productions are 
probably  essential. Nonetheless,    the presence of  a calculable deviation  from the sine formula is qualitatively  significant  and shows
 that such a ratio is not a universal  quantity.

  \section*{Acknowledgment}

The author thanks Misha Shifman and Misha Voloshin for providing us with a stimulating
atmosphere and occasions for fruitful discussions.


\begin{thebibliography}{100}
\footnotesize{

\bibitem{CKM}
G. Carlino, K. Konishi and H. Murayama,
   {\bf  JHEP  0002}  (2000) 004,     hep-th/0001036;
 {\bf    Nucl.Phys.  B590}  (2000) 37,     hep-th/0005076; 
K. Konishi, Proceedings of Continuous Advances in QCD, Minneapolis, Minnesota, May 2000,    hep-th/0006086;
G. Carlino, K. Konishi, Prem Kumar  and H. Murayama,
  {\bf    Nucl.Phys.  B608    }  (2001) 51, hep-th/0104064. 
 
\bibitem{ArPlSei}
P. C. Argyres, M. R. Plesser and N. Seiberg, {\bf Nucl. Phys.  B471}
(1996)
159, hep-th/9603042;
P.C. Argyres, M.R. Plesser, and A.D. Shapere,
 {\bf Nucl. Phys.  B483}   (1997) 172,   hep-th/9608129.
 


\bibitem{SW1}


N. Seiberg and E. Witten,   {\bf   Nucl.Phys. B426} (1994) 19; Erratum
\textit{ibid.}     {\bf   Nucl.Phys.   B430} (1994) 485, hep-th/9407087.

\bibitem{SW2}
N. Seiberg and E. Witten, {\bf Nucl. Phys.  B431} (1994) 484,
   hep-th/9408099.
   
\bibitem{curves}   

P.~C.~Argyres and A.~F.~Faraggi, {\bf Phys. Rev. Lett {\bf 74}} (1995)
3931, hep-th/9411047;
A. Klemm, W. Lerche, S. Theisen and S. Yankielowicz,  {\bf  Phys. Lett.
{\bf B344} }    (1995) 169, hep-th/9411048;
 {\bf  Int. J. Mod. Phys. A11}   (1996) 1929, hep-th/9505150;   
A. Hanany
and Y. Oz,   {\bf  Nucl. Phys. {\bf B452} }   (1995) 283,
hep-th/9505075;
P.  C.  Argyres, M.  R.  Plesser and A.  D.  Shapere,   {\bf  Phys.  Rev.
Lett.  {\bf 75} }      (1995) 1699, hep-th/9505100;
P. C. Argyres and A. D. Shapere,   {\bf    Nucl. Phys. {\bf B461} }    (1996) 437,
hep-th/9509175;   A. Hanany,
 {\bf   Nucl.Phys. {\bf B466}}    (1996) 85,  hep-th/9509176. 



\bibitem{TM}  G. 't Hooft, {\bf  Nucl. Phys.   B190}   (1981) 455.
   S. Mandelstam,  {\bf Phys. Lett.  53B }  (1975) 476;
                                {\bf  Phys. Rep.   23C} (1976) 245.



\bibitem{GNO}   P. Goddard, J. Nuyts and D. Olive,   {\bf Nucl. Phys.  B125}  (1977) 1.

\bibitem{EW}  E. Weinberg, {\bf Nucl. Phys. B167} (1980) 500;  {\bf Nucl. Phys. B203} (1982) 445. 

\bibitem{BK}  S. Bolognesi and K. Konishi, hep-th/0207161 (2002).






\bibitem{TP}    G. 't Hooft,  {\bf Nucl. Phys. B79} (1974) 276;  A.M. Polyakov, {\bf JETP Lett. 20}  (1974) 194.  




\bibitem{JR}  R. Jackiw and C. Rebbi, {\bf Phys. Rev.  D13} (1976) 3398; 
C. Callias, {\bf Comm. Math, Phys. 62} (1978) {213}; 
J. de Boer, K. Hori and Y. Oz, 
  {\bf Nucl. Phys.  B500} (1997) {163}, hep-th/9703100.


\bibitem{DPK}  M. Di Pierro and K. Konishi, {\bf Phys. Lett. B388} (1996) 90, 
 hep-th/9605178. 

\bibitem{Zum} 
 B. Zumino, Erice Lectures (1966), Ed. A. Zichichi; 
S. Coleman,  Erice Lectures (1977), Ed. A. Zichichi.

\bibitem{NO}   H. Nielsen and P. Olesen,   {\bf Nucl. Phys. B61}  (1973) 45.

\bibitem{DeVega}  H. J. de Vega,   {\bf  Phys. Rev.   D18 } (1978) 2932.



\bibitem{DS1}   H.J. de Vega and F.A. Schaposnik,  {\bf  Phys. Rev. Lett.  56} (1986) 2564;     
  {\bf  Phys. Rev.   D34 } (1986) 3206. 

\bibitem{HV}   J. Heo and T. Vachaspati,  {\bf  Phys. Rev.   D58}  (1998)
065011,   hep-ph/9801455. 

\bibitem{SS}  F.A. Schaposnik and P. Suranyi,  {\bf Phys. Rev. D62}  (2000) 125002,   hep-th/0005109. 

\bibitem{Kneipp}  M.A.C. Kneipp and P. Brockill, {\bf  Phys.Rev.D64}  (2001)  125012, hep-th/0104171.  


\bibitem{SY} M. Shifman and A. Yung, 
hep-th/0205025 (2002). 



\bibitem{STR}     A. Hanany, M. Strassler and A. Zaffaroni,  {\bf Nucl.Phys. B513}   (1998) 87,   hep-th/9707244.

\bibitem{DS}  M.R. Douglas and  S.H. Shenker,   {\bf Nucl. Phys.  B447}
(1995) 271,   hep-th/9503163.




\bibitem{LT}  B.  Lucini and  M. Teper,
{\bf Phys.Lett. B501} (2001) 128,   hep-lat/0012025;  
{\bf Phys.Rev.D64} (2001) 105019,
 hep-lat/0107007. 


\bibitem{Pisa} L. Del Debbio, H. Panagopoulos, P. Rossi and  E. Vicari,  {\bf Phys.Rev. D65} (2002) 021501, hep-th/0106185;
{\bf JHEP 0201} (2002) 009,  hep-th/0111090  


\bibitem{KI} C. P. Herzog and I. R. Klebanov,  {\bf Phys.Lett. B526} (2002) 388, 
hep-th/0111078. 

\bibitem{KA}  R. Auzzi and K. Konishi,   hep-th/0205172, NJP (2002), to appear. 

\bibitem{KSp} K. Konishi and L. Spanu, hep-th/0106075,  IJMPA  (2002), to appear. 

\bibitem{ABR}  A.A. Abrikosov, {\bf  JETP  5}  (1957) 1174.

\bibitem{STRASS}  M. Strassler,   {\bf Progr.  Theor. Phys. Suppl. 131} (1998) 439,  hep-lat/9803009.










\bibitem{YUNG}   A. Yung,   hep-th/0005088, 
3rd Moscow School of Physics and 28th ITEP Winter School of Physics, Moscow, 2000; 
 A. Vainshtein and A. Yung,  {\bf Nucl.Phys.B614},  3,2001,   hep-th/0012250.

 

  

\bibitem{HOU}
X.~r.~Hou,   {\bf Phys. Rev.  D  63} (2001) 045015, hep-th/0005119.

\bibitem{Fer}  F. Ferrari, {\bf Nucl.Phys. B612} (2001) 151,  hep-th/0106192 

\bibitem{Edel}  J.D. Edelstein, W.G. Fuertes,  J. Mas and  J. M. Guilarte, 
{\bf Phys.Rev. D62} (2000)  065008, hep-th/0001184. 

}
\end{thebibliography}
\end{document}